\documentstyle[prl,epsfig,aps]{revtex}
\include{psfig}

\begin{document}

\twocolumn[
 \hsize\textwidth\columnwidth\hsize
 \csname@twocolumnfalse\endcsname

\title{\bf Chaotic dot-superconductor analog of the Hanbury Brown
Twiss effect} 
\author{P. Samuelsson and M. B\"uttiker}
\address{D\'epartement de Physique Theorique, Universit\'e de
Gen\`eve, CH-1211 Gen\'eve 4, Switzerland.} \date{\today} \maketitle
\begin{abstract}
As an electrical analog of the optical Hanbury Brown Twiss effect, we
study current cross-correlations in a chaotic quantum
dot-superconductor junction. One superconducting and two normal
reservoirs are connected via point contacts to a chaotic quantum
dot. For a wide range of contact widths and transparencies, we find
large positive current correlations. The positive correlations are
generally enhanced by normal backscattering in the contacts. Moreover,
for normal backscattering in the contacts, the positive correlations
survive when suppressing the proximity effect in the dot with a weak
magnetic field. \\ \newline
\end{abstract}]

In quantum theory, identical particles are indistinguishable. Under
exchange of any pair of particles, the many-body wavefunction remains
invariant up to a sign, positive for bosons and negative for
fermions. This exchange symmetry was used in the pioneering
interferometer experiment with photons, by Hanbury Brown and Twiss
\cite{Hanbury56}. Several theoretical works
\cite{Buttiker90,Buttiker92,Buttikerrew} have suggested different
analogs of this experiment with electrons in mesoscopic multiterminal
conductors. It has been shown \cite{Buttiker92} that the fermionic
statistics of electrons leads to negative correlations between
currents flowing in different terminals. Such negative correlations
were also recently observed experimentally \cite{Oliver99}.

When normal conductors are connected to a superconductor, correlations
are introduced between electrons and holes due to Andreev reflections
at the normal-superconductor interface, a phenomenon known as the
proximity effect. The influence of the proximity effect on the current
auto-correlations, i.e. the shotnoise, in a two-terminal diffusive
normal-superconductor junctions was recently studied
\cite{Jehl00,Belzig01}.

In multiterminal conductors, Andreev reflection can lead to positive
cross correlations between currents flowing in the contacts to the
normal reservoirs\cite{Datta96,Torres99}. So far, positive
correlations have been predicted only for single mode junctions
\cite{Datta96,Torres99,Gramespacher99}. Moreover, in multiterminal
diffusive junctions it was found that cross correlations are negative
in the absence of the proximity effect\cite{Nagaev01}.

This raises two important questions: i) are the positive correlations
in normal-superconducting junctions a large effect, of the order of
the number of modes in multimode junctions, and if this is the case
ii) is the proximity effect necessary to obtain these positive
correlations? In this paper we give an answer to these two
questions. The positive correlations are large, and surprisingly, get
enhanced by normal backscattering at the normal-superconducting
interface. Moreover, positive correlations can exist even in the
absence of the proximity effect, if the normal-superconductor
interface is nonideal.

We study the current correlations in a system consisting of a chaotic
quantum dot connected via point contacts to one superconducting and
two normal reservoirs. Systems consisting of chaotic dots coupled to
superconductors have recently\cite{dotguys} attracted a lot of
interest. The generic properties of the model makes our result
qualitatively relevant for multiterminal normal-superconducting
structures with random scattering.
\begin{figure}[h]
\centerline{\psfig{figure=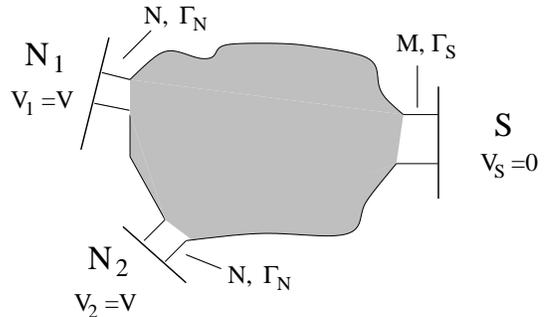,width=7.0cm}}
\caption{A chaotic quantum dot (grey shaded), acting as a
beam-splitter, is connected to two normal reservoirs ($N_1$ and $N_2$)
and one superconducting reservoir ($S$) via quantum point contacts.}
\label{fig1}
\end{figure}
A schematic picture of the system is shown in Fig. \ref{fig1}. A
quantum dot is connected to two normal reservoirs ($N_1$ and $N_2$)
and one superconducting reservoir ($S$) via quantum point contacts.
The contacts to the normal and superconducting reservoirs have mode
independent transparency $\Gamma_N$ and $\Gamma_S$ respectively and
support $N$ and $M$ transverse modes. The conductances of the point
contacts are much larger than the conductance quanta $2e^2/h$,
i.e. $N\Gamma_N,M\Gamma_S \gg 1$, so Coulomb blockade effects in the
dot can be neglected.  The two normal reservoirs are held at the same
potential $V$ and the potential of the superconducting reservoir is
zero.

We consider the case where the classical motion in the dot is chaotic
on time scales longer than the ergodic time $\tau_{erg}$.  The
quasiparticle dwell time in the dot, $\hbar/E_{Th}$, is assumed to be
much larger than $\tau_{erg}$, but much smaller than the inelastic
scattering time. Here $E_{Th}=(2N\Gamma_N+M\Gamma_S)\delta/\pi$, where
$\delta$ is the mean level spacing in the dot. Under these conditions
random matrix theory\cite{Beenakker97} can be used to describe the
statistical properties of the scattering matrix $S_d$ of the dot. The
scattering matrix $S$ of the combined system, dot and superconductor,
can be expressed \cite{Beenakker97} in terms of the scattering matrix
$S_d$ of the dot and the Andreev reflection amplitude at the
contact-superconductor interface.

Due to the random scattering in the dot, the current $I_i(t)$ in
contact $i$ fluctuates around its quantum statistical average $\bar
{I_i}$. We study the zero-frequency spectral density of the current
cross-correlations $P_{12}=2\int dt~\overline{\Delta I_1(t)\Delta
I_2(0)}$ where $\Delta I_j(t)=I_i(t)-\bar {I_i}$. The correlation can
be expressed in terms of the scattering matrix $S$. We consider the
limit of zero temperature and a potential $eV$ much lower than
$E_{Th}$ and $\Delta$, where the energy dependence of the scattering
matrix $S$ can be neglected. Moreover, at $|E|<\Delta$ no
quasiparticles can escape into the superconductor and the scattering
matrix $S$ describes only scattering between the normal reservoirs,
\begin{equation}
S=\left( \begin{array}{cc} S^{ee} & S^{eh} \\ S^{he} & S^{hh} \end{array}
\right),~S^{\alpha\beta}=\left( \begin{array}{cc} S^{\alpha\beta}_{11} &
S^{\alpha\beta}_{12} \\ S^{\alpha\beta}_{21} & S^{\alpha\beta}_{22}
\end{array} \right),
\label{stotmat}
\end{equation}
where $S^{\alpha\beta}_{ij}$ are matrix amplitudes ($N \times N$) for
injected quasiparticles (e or h) of type $\beta$ in lead $j$ to be back
reflected as quasiparticles of type $\alpha$ in lead $i$.

Noting that the current fluctuation is just the sum of the
fluctuations of electron and hole currents, the noise power can be
conveniently written \cite{Datta96},
\begin{equation}
P_{12}=P^{ee}_{12}+P^{hh}_{12}+P^{eh}_{12}+P^{he}_{12}
\label{totcorr}
\end{equation}
where the noise power $P_{12}^{\alpha\beta}$ for correlation between
quasiparticle currents, is given
by
\begin{equation}
P_{12}^{\alpha\beta}=\sigma V\frac{4e^3}{h}\sum_{i,j=1,2}\mbox{tr}\left[(S^{\alpha
e}_{1i})^{\dagger}S^{\alpha h}_{1j}(S^{\beta h}_{2j})^{\dagger}S^{\beta
e}_{2i}\right],
\label{qpnoise}
\end{equation}
with $\sigma=+(-)$ for $\alpha=\beta~(\alpha \neq \beta)$. The
expression for the correlations in
Eqs. (\ref{totcorr})-(\ref{qpnoise}) is an extension of the result
\cite{Buttiker92} for a purely normal conductor, taking into account
both electron and hole quasiparticles. However, unlike the normal
cross correlations, which are manifestly negative, the
cross-correlation $P_{12}$, can be positive, because the correlations
between different types of quasiparticles, $P^{eh}_{12}+P^{he}_{12}$,
are positive. Note however that
$P_{12}^{ee}+P_{12}^{hh}-(P_{12}^{eh}+P_{12}^{he})$, the
cross-correlator between probability currents, is manifestly
negative\cite{Datta96}, a consequence of the ferminonic statistics of
the quasiparticles.

The ensemble averaged correlations $\langle P_{12} \rangle$ are
calculated using the statistical properties of the scattering matrix
$S_d$ of the dot. We first consider the case with ideal contacts
$\Gamma_N,\Gamma_S=1$ and no magnetic field in the dot, i.e. conserved
time reversal symmetry. In this case it is useful to decompose $S_d$
as (see e.g. Ref.\cite{Beenakker97})
\begin{equation}
S_d=\left(\begin{array}{cc} U & 0 \\ 0 & U' \end{array}
\right)\left(\begin{array}{cc} r_{2N,2N} & t_{2N,M} \\ t_{M,2N} &
r_{M,M} \end{array} \right)\left(\begin{array}{cc} U^T & 0 \\ 0 &
U'^T \end{array} \right),
\label{poldecomp}
\end{equation}
where $U$($U'$) is a unitary matrix of dimension $2N \times 2N$
($M\times M$), uniformly distributed in the unitary group. The
diagonal matrices $r_{2N,2N}$ and $r_{M,M}$ have $\mbox{min}(2N,M)$
elements $\sqrt{1-T_n}$ and the rest unity ($t_{2N,M}$ and $t_{M,2N}$
follow from the unitarity of $S_d$). Here $T_n$ are the transmission
eigenvalues, which have a density\cite{Nazarov95}
$\rho(T)=(2N+M)/(2\pi)\sqrt{T-T_{min}}/(T\sqrt{1-T})$, where
$T_{min}=[(2N-M)/(2N+M)]^2$ is the smallest possible eigenvalue.

\begin{figure}[h]
\centerline{\psfig{figure=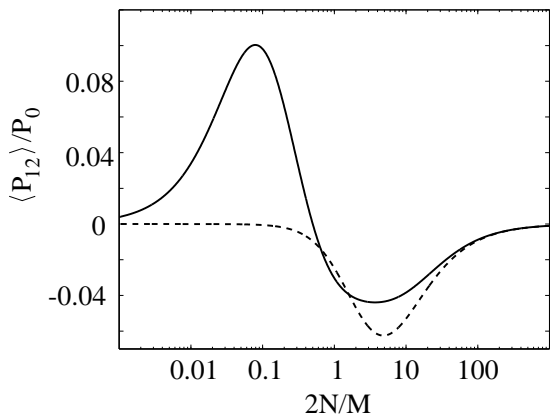,width=7.5cm}}
\vspace{0.4cm}
\caption{The ensemble averaged current cross-correlation $\langle
P_{12} \rangle$ with (solid) and without (dashed) proximity effect in
the dot, as a function of $\gamma=2N/M$. The correlations in the
presence of the proximity effect are positive for a dominating
coupling to the superconductor ($\gamma <0.5$), but are negative for
all $\gamma$ when suppressing the proximity effect with a weak
magnetic field in the dot.}
\label{fig2}
\end{figure}

Inserting the decomposition into the expression for the total
scattering matrix $S$ we find \cite{Beenakker97} the quasiparticle
scattering amplitudes $S^{\alpha\beta}$ in Eq. (\ref{stotmat}) as e.g
$S^{ee}=U\sqrt{1-T}/(2-T)U^*$ ($T=1-r_{2N,2N}^2$). Summing over
injection contacts in the expression for the individual quasiparticle
current correlators in Eq. (\ref{qpnoise}) and inserting the
scattering amplitudes we get
\begin{eqnarray}
P_{12}^{ee}=V\frac{4e^3}{h}\mbox{tr}[U(1-{\cal R})U^*C_1U^T{\cal
R}U^{\dagger}C_2]
\label{qpnoise2}
\end{eqnarray}
and similar for the other terms $P^{\alpha\beta}_{12}$. Here ${\cal
R}=T^2/(2-T)^2$ are eigenvalues of the matrix product
$(S^{eh})^{\dagger}S^{eh}$ and the matrix $C_1$ is diagonal with
elements $(C_1)_n=1$ for $n\leq N$ and $0$ otherwise. The matrix
$C_2=1-C_1$. The ensemble average of Eq. (\ref{qpnoise2}) is
carried out in two steps. First, $P^{\alpha\beta}_{12}$ is averaged
over the unitary matrix $U$, using the diagrammatic technique in
Ref. \onlinecite{Brouwer97}. This gives to leading order in $N,M$ (i.e
neglecting weak localization corrections) for the total current
corellations in Eq. (\ref{totcorr}),
\begin{equation}
\langle P_{12}\rangle_U=V\frac{4e^3}{h}\left(\mbox{tr}[{\cal
R}(1-{\cal R})]-\frac{\mbox{tr}{(\cal R})\mbox{tr}(1-{\cal
R})}{4N}\right).
\label{transeig}
\end{equation}
We then perform the average over transmission eigenvalues by
integrating $\langle P_{12} \rangle_U$ weighted by the density
$\rho(T)$. Using that to leading order in $N$ we have
$\langle\mbox{tr}({\cal R})\mbox{tr}(1-{\cal
R})\rangle_T=\langle\mbox{tr}({\cal R})\rangle_T
\langle\mbox{tr}(1-{\cal R})\rangle_T$, we get ($<..>=<..>_{U,T}$)
\begin{equation}
\frac{\langle P_{12} \rangle}{P_0}=\frac{(1+\gamma)^2}{2\gamma^2}\left(\frac{1}{1+\gamma}-2\frac{\gamma}{f}-\frac{f^2-16\gamma^3}{f^{5/2}}\right)
\end{equation}
where $f(\gamma)=1+6\gamma+\gamma^2$ and $\gamma=2N/M$, the ratio
between the conductances of the point contacts connected to the normal
and the superconducting reservoirs. The correlation is normalized with
$P_{0}=4VNe^3/h$, i.e. it is large, of order $N$. The correlation
$\langle P_{12} \rangle$, plotted in Fig. \ref{fig2}, is positive for
a dominating coupling to the superconductor, but crosses over (at
$\gamma\approx 0.5$) to negative values upon increasing the coupling
to the normal reservoirs.

In the general case, it is difficult to provide a simple explanation
for the sign and magnitude of the correlations $\langle P_{12}
\rangle$, since they result from a competition between $\langle
P_{12}^{ee} \rangle+\langle P_{12}^{hh} \rangle$ and $\langle
P_{12}^{eh} \rangle+\langle P_{12}^{he} \rangle$, which have opposite
sign and generally are of the same magnitude. It is however possible
to get a qualitative picture in certain limits by studying the
different contributing scattering processes.
In the limit $\gamma \rightarrow 0$, the coupling to the normal
reservoirs is negligible and a gap in the spectrum opens up around
Fermi energy in the dot\cite{dotguys}. As a consequence,
quasiparticles injected from one normal reservoir are Andreev
reflected, effectively direct at the contact-dot
interface\cite{Clerk00}, with unity probability back to the same
reservoir. There is thus no partition of incoming quasiparticles and
hence no noise, $P_{12}=0$.

Increasing the coupling to the normal reservoir, the probability of
normal reflection($\sim \gamma$) as well as cross Andreev reflection
($\sim \gamma^2$), from one reservoir to the other, becomes
finite. Since normal reflection is the dominant process, we can
neglect the terms in $P_{12}$ in Eq. (\ref{qpnoise}) containing the
cross Andreev reflection amplitude (e.g. $S_{12}^{eh}$). This gives
$\langle P_{12}\rangle=8V(e^3/h)\langle
\mbox{tr}(S^{ee}_{12}(S^{ee}_{12})^{\dagger}S^{eh}_{11}(S^{eh}_{11})^{\dagger})\rangle=2P_0\gamma$,
positive since only terms in $\langle P_{12}^{eh}\rangle+\langle
P_{12}^{he}\rangle$ contribute. This expression shows that the
correlations can be explained as partition noise of injected
electrons, which have probability $\lesssim 1$ to be Andreev reflected
(effectively at the contact-dot interface) and probability $\sim
\gamma$ to be normally reflected.
 
In the opposite limit, $\gamma \gg 1$, the coupling to the
superconductor is weak and the Andreev reflection probability is
small, of order $1/\gamma$. The correlation can be written as $\langle
P_{12} \rangle=-16V(e^3/h)\langle
\mbox{tr}(S^{ee}_{11}(S^{ee}_{11})^{\dagger}S^{eh}_{21}(S^{eh}_{21})^{\dagger})\rangle=-P_0/\gamma$,
which is negative since we find that in this limit only terms in
$\langle P_{12}^{ee} \rangle+\langle P_{12}^{hh} \rangle$
contribute. This shows that the correlation can be explained as
partition noise of electrons which have probability $\lesssim 1/2$ to
be normally reflected (without reaching the dot-superconductor
contact) and probability $\sim 1/\gamma$ to be Andreev reflected.

When the proximity effect is suppressed by a weak magnetic field in
the dot, \cite{Clerk00} it is no longer possible to express the
correlations directly in terms of the transmission eigenvalues $T$, as
in Eq. (\ref{transeig}). We instead use the fact that the scattering
matrix $S_d$ of the dot itself is uniformly distributed in the unitary
group, without constraints on symmetry of $S_d$. Since the scattering
matrix amplitudes $S^{\alpha\beta}$ in Eq. (\ref{qpnoise}) can be
expressed \cite{Beenakker97} in terms of $S_d$, the correlations
$P_{12}^{\alpha\beta}$ can be directly averaged over $S_d$ using the
diagrammatic technique in Ref. \onlinecite{Brouwer97}. This gives for
the total correlations
\begin{equation}
\frac {\langle P_{12}
\rangle}{P_0}=-\frac{\gamma^2(1+\gamma)}{(2+\gamma)^4},
\end{equation}
plotted in Fig. \ref{fig2}. The gap in the spectrum is suppressed and
the scattering processes responsible for the positive correlations in
the presence of the proximity effect are strongly modified, most
notably in the limit $\gamma \ll 1$. As a consequence, the
correlations are manifestly negative for all $\gamma$, i.e. there are
no positive correlations in the absence of the proximity effect,
similar to what was found for a metallic diffusive system
\cite{Nagaev01}.

Until now we have only considered ideal point contacts, with
$\Gamma_N,\Gamma_S=1$. In an experimental situation, it is often
difficult to obtain an ideal contact between the dot and the
superconductor. It is therefore of interest to study the situation
with a nonideal dot-superconductor contact. We consider first the case
with proximity effect in the dot. To calculate the current
correlations in this case we note \cite{Beenakker97} that a nonideal
interface changes the density of transmission eigenvalues, $\rho(T)$,
but not the distributions of the unitary matrix $U$ in
Eq. (\ref{qpnoise2}). The transmission eigenvalue density is
calculated numerically \cite{Nazarov95,Brouwer97} for different
contact transparencies $\Gamma_S<1$ and the integrals in
Eq. (\ref{transeig}) are then performed.

\begin{figure}[h]
\centerline{\psfig{figure=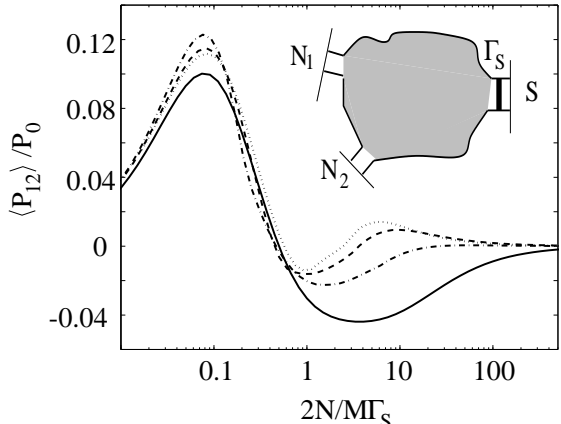,width=7.5cm}}
\caption{Ensemble averaged current cross correlations $P_{12}$ in the
presence of normal backscattering at the normal-superconducting
interface (see inset). The contact transparencies $\Gamma_S$ are $1$
(solid), $0.8$ (dash-dotted) $0.6$ (dashed) and $0.4$ (dotted). The
correlations are plotted as a function of $2N/(M\Gamma_S)$. Note that
the correlations for a dominating coupling to the normal reservoirs
crosses over from negative to positive on {\it increasing} the normal
backscattering.}
\label{fig3}
\end{figure}

The resulting correlations are plotted in
Fig. \ref{fig3}. Surprisingly, the main effect of normal
backscattering at the dot-superconductor contact is to cause a
crossover from negative to positive correlation for a strong coupling
to the normal reservoirs. In this limit, $\gamma \gg 1$,
injected quasiparticles undergo at the most one scattering event at
the dot-superconductor contact before leaving the junction. The
expression for the scattering matrices $S_{ij}^{\alpha\beta}$ in
Eq. (\ref{qpnoise}) simplify considerably and we can derive an
analytical expression for the correlations, giving\cite{torrescom}
\begin{equation}
\frac{\langle P_{12}\rangle}{P_0}=\frac{1}{\gamma}R_{eh}(1-2R_{eh})
\label{notrssmallar}
\end{equation}
where $R_{eh}=\Gamma_S^2/(2-\Gamma_S)^2$ is the Andreev reflection
probability of quasiparticles incident in the dot-superconductor
contact. There is a crossover from negative to positive correlations
already for $R_{eh}=1/2$, i.e $\Gamma_S=2(\sqrt{2}-1)\approx 0.83$, in
agreement with the full numerics in Fig. \ref{fig3}.

Since $\Gamma_S<1$, now $R_{eh}$ is smaller than one. As a
consequence, there is additional partition due to the possibility of
normal reflection at the dot-superconductor contact. The partition
noise discussed above, from electrons being either Andreev reflected
or normally reflected (without reaching the dot-superconductor
contact), is thus reduced to $-P_0R_{eh}/\gamma$. However, the
additional normal reflection at the dot-superconductor contact give
rise to a noise term $2P_0R_{eh}(1-R_{eh})/\gamma$, with opposite sign
[together they give Eq. (\ref{notrssmallar})]. For $R_{eh}<1/2$, the
second term is dominating, causing the crossover to positive
correlations.
Interestingly, in the absence of a proximity effect in the dot, we
find in the same way that the correlation for $\gamma \gg 1$ is also
given by Eq. (\ref{notrssmallar}), the argument being the same as in
the presence of the proximity effect. The proximity effect thus plays
no role in this limit, where the quasiparticles undergo at most one
Andreev reflection. This shows that the proximity effect is not a
necessary condition for positive correlations.

Finally, we note that the effect of normal backscattering in the
contacts between the dot and the normal reservoirs is to enhance the
positive correlations for a dominating coupling to the superconducting
reservoir. In the limit with tunnel barriers in all contacts, the
transmission eigenvalue density is known analytically\cite{Nazarov95}
and the correlation $\langle P_{12} \rangle$ follows from
Eq. (\ref{transeig}) \cite{Belzigcom}
\begin{equation}
\frac{\langle P_{12}\rangle}{P_0\Gamma_N}=\frac{{\bar
\gamma}}{(1+{\bar \gamma}^2)^{3/2}}\left(1-5\frac{{\bar
\gamma}^2}{(1+{\bar \gamma}^2)^{2}}\right),
\label{tunnellim}
\end{equation}
where ${\bar \gamma}=2N\Gamma_N/(M\Gamma_S)$. The correlations are
thus positive for ${\bar \gamma} <(\sqrt{5}-1)/2$ and ${\bar
\gamma}>(\sqrt{5}+1)/2$, and negative for intermediate values.

In conclusion, we have studied the current cross-correlations in a
three terminal superconducting-chaotic dot analog of the Hanbury Brown
Twiss interferometer. We find that the correlations are positive for a
wide range of junction parameters, and survive even in the absence of
a proximity effect in the dot. The magnitude of the positive
correlations is large, proportional to the number of transport modes
in the contacts to the dot, which should simplify an experimental
observation.

We acknowledge discussions with C.W.J. Beenakker, W. Belzig,
E. Sukhorukov, and G. Johansson. The work was supported by the Swiss
National Science foundation and the program for Materials with Novel
Electronic Properties. In the final stages of this work we became
aware of Boerlin et al. \onlinecite{Belzig}, who investigate a three
terminal junction in the tunnel limit. In this limit, our results
coincide.
%


\begin{thebibliography}{999}
\bibitem{torrescom} 
We note that this result coincides with what is
found for the single mode model of Ref. \cite{Torres99}, upon
substituting $\Gamma_S=2\epsilon$.
\bibitem{Belzigcom} 
The same result was obtained in Ref. \cite{Belzig}, see
comment at the end of the paper.
\end{thebibliography}
\end{document}